\documentclass[12pt]{article}

\renewcommand{\title}[1]{
\begin{center} \Large \bf #1 \end{center}
}

\renewcommand{\author}[2]{
 \begin{center} #1  \vspace{3mm} \\
  #2 \\
 \end{center}
\addvspace{\baselineskip}
}

\usepackage{amssymb}
\usepackage{amsmath}

\usepackage{amsthm}
\newtheorem{thm}{Theorem}[section]
\newtheorem{prop}[thm]{Proposition}

\theoremstyle{definition}
\newtheorem{defn}{Definition}

\theoremstyle{remark}

\makeatletter
\@addtoreset{equation}{section}
\def\theequation{\thesection.\arabic{equation}}
\makeatother
\begin{document}

\baselineskip 5mm

\title{
Explicit Formulae for Noncommutative Deformations of 
${\mathbb C}\mathbf{P^N}$ 
and ${\mathbb C}\mathbf{H^N}$}

\author{Akifumi Sako, Toshiya Suzuki and~ Hiroshi Umetsu }{
Kushiro National College of Technology\\
Otanoshike-Nishi 2-32-1, Kushiro 084-0916, Japan }

\vspace{1cm}

\abstract{ We give explicit expressions of a deformation quantization
with separation of variables for ${\mathbb C}P^N$ and ${\mathbb C}H^N$.
This quantization method is one of the ways to perform a deformation
quantization of K\"ahler manifolds, which is introduced by Karabegov.
Star products are obtained as explicit formulae in all order in the
noncommutative parameter.  We also give the Fock representations of 
the noncommutative ${\mathbb C}P^N$ and ${\mathbb C}H^N$.}

%\newpage

\section{Introduction}

Deformation quantizations were introduced by Bayen, Flato, Fronsdal,
Lichnerowicz and Sternheimer \cite{Bayen:1977ha} as a method to quantize
spaces. After \cite{Bayen:1977ha}, several ways of deformation
quantization were proposed \cite{DeW-Lec, Omori, Fedosov, Kontsevich}.
In particular, deformation quantizations of K\"ahler manifolds were
provided in \cite{Moreno86a, Moreno86b, Cahen93, Cahen95}\footnote{For a
recent review, see \cite{Schlichenmaier}.}. 
In this article, we consider the deformation
quantization with separation of variables that are one of the ways to
construct the noncommutative K\"ahler manifolds introduced by Karabegov
\cite{Karabegov, Karabegov1996, Karabegov2011}.

In many cases, deformation quantization of a manifold is given by a
star product which is defined in a form of a formal power series of
deformation parameter $\hbar$.  The power series is obtained as
solutions of an infinite system of differential equations, and it is
proved that there exists a unique deformation quantization as the
solution of the system.  The existence of the solution is proved for a
wide class of manifolds, however explicit expressions of deformation
quantizations are constructed only for few kinds of manifolds.  For
example, Euclidean spaces are deformed by using the Moyal product, and
on manifolds with spherically symmetric metrics explicit star products
are given in the context of the Fedosov's deformation quantization
\cite{Fedosov}.

The aim of this article is to give explicit expressions of 
deformation quantizations with separation of variables for ${\mathbb
C}P^N$ and ${\mathbb C}H^N$.\footnote{Star products on the fuzzy
${\mathbb C}P^N$ are investigated in \cite{Balachandran, Kitazawa,
Karabali}. A deformation quantization of the hyperbolic plane was
provided in \cite{Bieliavsky}.}  To construct star products, we have
to solve the infinite system of differential equations. In these cases,
as will be shown, differential equation systems are solvable, and 
expressions of star products are explicitly given in all order of
$\hbar$. A noncommutative deformation of $\mathbb{C}P^N$ was
investigated by performing the phase space reduction in
\cite{Bordemann}. We will comment on the connection between their star
product and our result.

We also give the Fock representations of noncommutative 
${\mathbb C}P^N$ and ${\mathbb C}H^N$.
The Fock representations of star products are also used in
investigations of field theories on noncommutative spaces. In
particular, the Fock representations give useful methods of
constructing solitons and instantons in noncommutative field
theories. Further, matrix models corresponding to noncommutative field
theories can be obtained from the Fock representations, and
quantum analyses of the models are actively pursued.

The organization of this article is as follows.
In Section \ref{Kahler}, we review the deformation quantization
with separation of variables proposed by Karabegov.
In Section \ref{CPN}, a star product for ${\mathbb C}P^N$
is given explicitly by using the deformation quantization
with separation of variables.
In Section \ref{FOCK}, we give the Fock representation of 
the star product obtained in Section \ref{CPN}.
In Section \ref{CHN}, a star product for ${\mathbb C}H^N$
is constructed explicitly by the similar way to 
the one in ${\mathbb C}P^N$.
Finally, we summarize our results and discuss their several perspectives
in Section \ref{Summary}.

%%%%%%%%%%%%%%%%%%%%%%%%%%%%%%%%%%%%%%%%%%%%%%%%%%%%%%%%%%%%%%%%
\section{Review of the deformation quantization with separation of
 variables} 
\label{Kahler}
%%%%%%%%%%%%%%%%%%%%%%%%%%%%%%%%%%%%%%%%%%%%%%%%%%%%%%%%%%%%%%%%%%%%%%%%%%%

In this section, we review the deformation quantization with
separation of variables 
to construct noncommutative K\"ahler manifolds.

An $N$-dimensional complex K\"ahler manifolds is defined by 
using a K\"ahler potential.
Let $\Phi$ be a K{\" a}hler potential and $\omega$
be a K{\" a}hler 2-form:
\begin{eqnarray}
\omega &:=& i g_{k \bar{l}} dz^{k} \wedge d \bar{z}^{l} ,
\nonumber \\
g_{k \bar{l}} &:=& 
\frac{\partial^2 \Phi}{\partial z^{k} \partial \bar{z}^{l}} .
\end{eqnarray}
In this paper, we use the Einstein summation convention over repeated
indices. 
The $g^{\bar{k} l}$ is the inverse of the metric $g_{k \bar{l}}$:
\begin{eqnarray}
 g^{\bar{k} l}  g_{l \bar{m}} = \delta_{\bar{k} \bar{m} } .
\end{eqnarray}
In the following, we denote 
\begin{align}
\partial_k = \frac{\partial}{\partial z^{k}} , ~
\partial_{\bar{k}} = \frac{\partial}{\partial \bar{z}^{k}}.
\end{align}

Deformation quantization is defined as follows.
\begin{defn}[Deformation quantization (weak sense)]
Deformation quantization is defined as follows.
$\cal F$ is defined as a set of formal power series:
\begin{eqnarray}
{\cal F} := \left\{  f \ \Big| \ 
f = \sum_k f_k \hbar^k, ~f_k \in C^\infty 
\right\} .
\end{eqnarray}
A star product is defined as 
\begin{eqnarray}
f * g = \sum_k C_k (f,g) \hbar^k
\end{eqnarray}
such that the product satisfies the following conditions.
\begin{enumerate}
\item $*$ is associative product.
\item $C_k$ is a bidifferential operator.
\item $C_0$ and $C_1$ are defined as 
\begin{eqnarray}
&& C_0 (f,g) = f g,  \\
&&C_1(f,g)-C_1(g,f) = i \{ f, g \}, \label{weakdeformation}
\end{eqnarray}
where $\{ f, g \}$ is the Poisson bracket.
\item $ f * 1 = 1* f = f$.
\end{enumerate}
\end{defn}
Note that this definition of the deformation 
quantization is weaker than the usual definition
of deformation quantization.
The difference between them is in (\ref{weakdeformation}).
In the strong sense of deformation quantization the condition 
$C_1(f,g)= \frac{i}{2} \{ f, g \}$ 
is required. 
For example, the Moyal product satisfies this condition.
But deformation quantizations with the separation of variables do
not satisfy this condition. 
In the following, ``deformation quantization'' is
used in this weak sense.

\begin{defn}[A star product with separation of variables]
$*$ is called a star product with separation of variables when 
\begin{eqnarray}
a * f = a f 
\end{eqnarray}
for a holomorphic function $a$ and
\begin{eqnarray} 
f * b = f b
\end{eqnarray}
for an anti-holomorphic function $b$.
\end{defn}

We use 
$$
D^{\bar{l}} = g^{\bar{l} k } \partial_k
= i\{ \bar{z}^l , \cdot \}
$$
and 
$$
{\cal S} := 
\Big\{ 
A | A=\sum_{\alpha} a_{\alpha} D^{\alpha} ,\ \ 
a_{\alpha} \in C^{\infty}
\Big\} ,
$$
where ${\alpha}$ is a multi-index 
$\alpha = (\alpha_1 , \alpha_2 , \dots , \alpha_n)$.

There are some useful formulae.
$D^{\bar{l}}$ satisfies the following equations.
\begin{align}
[ D^{\bar{l}} , D^{\bar{m}} ] &= 0 \ \ ,\ \   \forall l,m  \label{lem1-1}\\
[ D^{\bar{l}} , \partial_{\bar{m}} \Phi ] 
&= {\delta^{\bar{l}}}_{\bar{m}} \label{lem1-2}\\
\partial_k = g_{k \bar{l}} D^{\bar{l}}  . \label{lem1-3}
\end{align}

Using them, one can construct
a star product as differential operator $A_f$ such that
$f*g=A_f g$.
\begin{thm} \label{theo1}
For arbitrary $\omega$, there exist a star product with separation of
variables $*$ and it is constructed as follows.  Let $f$ be an element
of ${\cal F}$ and $A_n \in {\cal S}$ be a differential operator whose
coefficients depend on $f$ i.e.
\begin{eqnarray}
A_n = a_{n, \alpha}(f) D^{\alpha} , \ 
D^{\alpha}= \prod_{i=1}^n (D^{\bar{i}})^{\alpha_i} , \ 
(D^{\bar{i}}) =g^{\bar{i} l} \partial_l ,
\end{eqnarray}
where %$\displaystyle \partial_l= \frac{\partial}{\partial z^l}$,
$\alpha$ is an multi-index $\alpha = (\alpha_1 , \alpha_2 , \dots , \alpha_n)$.
Then,
\begin{eqnarray}
\tilde{A}_f = \sum_{n=0}^{\infty} \hbar^n A_n 
\end{eqnarray}
is uniquely determined such that
it satisfies the following conditions.
\begin{enumerate}
\item When $\displaystyle R_{\partial_{\bar{l}} \Phi} = 
\partial_{\bar{l}}\Phi + \hbar \partial_{\bar{l}}$,
\begin{eqnarray}
\left[  \tilde{A}_f , R_{\partial_{\bar{l}} \Phi} \right]=0 \ .
\label{A_f_R_Phi}
\end{eqnarray}
\item 
\begin{eqnarray}
\tilde{A}_f 1 &=& f*1=f , \label{A_f_2_1}\\
\tilde{A}_f g &:=& f * g , \label{A_f_2_2}\\
\tilde{A}_h ( \tilde{A}_g f ) &=& h * (g * f)
= (h*g)*f = \tilde{A}_{\tilde{A}_h g} f . \label{A_f_2_3}
\end{eqnarray}
\end{enumerate}
\end{thm}
Recall that each two of $D^{\bar{i}}$ 
commute each other, so if multi index $\alpha$ is fixed
then the $A_n$ is uniquely determined.
These conditions (\ref{A_f_2_1})-(\ref{A_f_2_3})
teach us that $\tilde{A}_f g =f * g$
is deformation quantization.

The following proposition is used in Section 3 and Section 5.
\begin{prop} \label{prop1}
We denote a left operation for a generic function 
$f$ as $L_f := \tilde{A}_f $ i.e. $L_f g =f* g$.
The right operation for $f$ is defined similarly
by $R_f g := g * f$.
$L_f $ ($R_f$) is obtained by using $L_{\bar{z}^l}$ ($R_{z^l}$) where
$L_{\bar{z}^l}$ ($R_{z^l}$) is defined by 
$L_{\bar{z}^l} g = \bar{z}^l * g$ ($R_{z^l} g = g*z^l$):
\begin{align}
L_f &= \sum_{\alpha} \frac{1}{\alpha !} 
\left( \frac{\partial}{\partial \bar{z}} \right)^{\alpha} f \ 
(L_{\bar{z}} -\bar{z} )^{\alpha}, \label{L_f} \\
R_f &= \sum_{\alpha} \frac{1}{\alpha !} 
\left( \frac{\partial}{\partial {z}} \right)^{\alpha} f \ 
(R_{{z}} -{z} )^{\alpha}. \label{R_f}
\end{align}
\end{prop}

%%%%%%%%%%%%%%%%%%%%%%%%%%%%%%%%%%%%%%%%%%%%
\section{Star product with separation of variables on
 $\mathbb{C}\mathbf{P^N}$}
 \label{CPN}
%%%%%%%%%%%%%%%%%%%%%%%%%%
In the inhomogeneous coordinates $z^i ~(i=1, 2, \cdots, N)$, the K\"ahler
potential of $\mathbb{C}P^N$ is given by
\begin{align}
 \Phi = \ln \left(1+|z|^2\right), \label{phi}
\end{align}
where $|z|^2 = \sum_{k=1}^N z^k \bar{z}^k$.
The metric $(g_{i\bar{j}})$ is 
\begin{align}
 ds^2 &= 2g_{i\bar{j}}dz^id\bar{z}^j, \label{ds} \\
 g_{i\bar{j}} &= \partial_i \partial_{\bar{j}} \Phi
  = \frac{(1+|z|^2)\delta_{ij}-z^j \bar{z}^i}{(1+|z|^2)^2}, \label{metric}
\end{align}
and the inverse of the metric $(g^{\bar{i}j})$ is
\begin{align}
 g^{\bar{i}j} = (1+|z|^2)\left(\delta_{ij}+z^j\bar{z}^i\right). \label{inverse}
\end{align}
%We use the notations, 
%$\partial_i = \partial/\partial z^i, 
%\partial_{\bar{i}} = \partial/\partial \bar{z}^i$.
%Define the differential operators $D^i$ and $D^{\bar{i}}$ as
%\begin{align}
% D^i = g^{i\bar{j}}\partial_{\bar{j}}, \qquad
%  D^{\bar{i}} = g^{\bar{i}j}\partial_j.
%\end{align} 
%In this paper, we use the Einstein summation convention over repeated
%indices. 

We here summarize useful relations in the following calculations;
\begin{align}
 \partial_{\bar{i}_1} \partial_{\bar{i}_2} \cdots 
  \partial_{\bar{i}_n} \Phi 
  = &~ (-1)^{n-1} (n-1)! ~\partial_{\bar{i}_1}\Phi
  \partial_{\bar{i}_2}\Phi \cdots \partial_{\bar{i}_n}\Phi,
 \label{par-phi} \\
 \left[\partial_{\bar{i}}, D^{\bar{j}}\right]
  =&~ \partial_{\bar{i}}\Phi D^{\bar{j}} 
  + \delta_{ij}\partial_{\bar{k}}\Phi D^{\bar{k}}, \label{par-D} \\
 \left[\partial_{\bar{i}}, ~c_{\bar{j}_1 \bar{j}_2 \cdots \bar{j}_n}
 D^{\bar{j}_1} D^{\bar{j}_2} \cdots D^{\bar{j}_n} \right]
  =&~ \partial_{\bar{i}} c_{\bar{j}_1 \bar{j}_2 \cdots \bar{j}_n}
 D^{\bar{j}_1} D^{\bar{j}_2} \cdots D^{\bar{j}_n}
 + n c_{\bar{j}_1 \cdots \bar{j}_n} \partial_{\bar{i}}\Phi 
 D^{\bar{j}_1} \cdots D^{\bar{j}_n} \nonumber \\
 & \hspace{-30mm} 
 + n c_{\bar{i}\bar{j}_1 \cdots \bar{j}_{n-1}} \partial_{\bar{k}} \Phi
  D^{\bar{k}} D^{\bar{j}_1} \cdots D^{\bar{j}_{n-1}} 
%  \nonumber \\
% &~ 
 + n(n-1) c_{\bar{i} \bar{j}_1 \cdots \bar{j}_{n-1}}
  D^{\bar{j}_1} \cdots D^{\bar{j}_{n-1}}, \label{par-a-D}
\end{align}
where the coefficients $c_{\bar{j}_1 \bar{j}_2 \cdots \bar{j}_n}$ are
totally symmetric under the permutations of the indices.

We construct the operator $L_{\bar{z}^l}$, which is corresponding to the
left star product by $\bar{z}^l$. $L_{\bar{z}^l}$ is defined as a
power series of $\hbar$, 
\begin{align}
 L_{\bar{z}^l} = \bar{z}^l + \hbar D^{\bar{l}}
  + \sum_{n=2}^\infty \hbar^n A_n, \label{Lzh}
\end{align}
where $A_n ~(n\geq 2)$ is a formal series of the differential operators
$D^{\bar{k}}$. We assume that $A_n$ has the following form,  
\begin{align}
 \label{An}
 A_n = \sum_{m=2}^n a^{(n)}_m
  \partial_{\bar{j}_1}\Phi \cdots \partial_{\bar{j}_{m-1}}\Phi
  D^{\bar{j}_1} \cdots D^{\bar{j}_{m-1}} D^{\bar{l}},
\end{align}
where the coefficients $a^{(n)}_m$ do not depend on $z^i$ and $\bar{z}^i$.

{}From the requirement of $\left[L_{\bar{z}^l}, \partial_{\bar{i}}\Phi +
\hbar \partial_{\bar{i}} \right] = 0$, the operators $A_n$ are recursively
determined by the equations
\begin{align}
 \label{rec-An}
 \left[ A_n, \partial_{\bar{i}}\Phi \right]
  = \left[ \partial_{\bar{i}}, A_{n-1}\right], \qquad (n\geq 2)
\end{align}
where $A_1 = D^{\bar{l}}$.  
$A_2 = \partial_{\bar{j}}\Phi D^{\bar{j}} D^{\bar{l}}$ is easily
obtained from the above equation.
Using the expression (\ref{An}), the left hand side of the recursion
relation (\ref{rec-An}) becomes 
\begin{align}
 \left[ A_n, \partial_{\bar{i}}\Phi \right]
  &= \sum_{m=2}^n a^{(n)}_m
  \partial_{\bar{j}_1}\Phi \cdots \partial_{\bar{j}_{m-1}}\Phi
  \left[ D^{\bar{j}_1} \cdots D^{\bar{j}_m} D^{\bar{l}}, 
   \partial_{\bar{i}}\Phi \right]
  \nonumber \\
% &= \sum_{m=2}^n m c^{(n)}_m 
%  \left\{ 
%  (m-1) \partial_{\bar{j}_1}\Phi \cdots \partial_{\bar{j}_{m-2}}\Phi
%  \partial_{\bar{i}}\Phi
%  D^{\bar{j}_1} \cdots D^{\bar{j}_{m-2}} D^{\bar{l}}
%  \right. \nonumber \\
%  && \left. ~~~~+ \delta_{il} \partial_{\bar{j}_1}\Phi \cdots
%  \partial_{\bar{j}_{m-1}}\Phi
%  D^{\bar{j}_1} \cdots D^{\bar{j}_{m-1}}
%  \right\}
%  \nonumber \\
 &= \sum_{m=2}^{n-1} a^{(n)}_{m+1}
  \left\{
  m \partial_{\bar{j}_1}\Phi \cdots \partial_{\bar{j}_{m-1}}\Phi
  \partial_{\bar{i}}\Phi
  D^{\bar{j}_1} \cdots D^{\bar{j}_{m-1}} D^{\bar{l}}
 + \delta_{il} \partial_{\bar{j}_1}\Phi \cdots
     \partial_{\bar{j}_m}\Phi
  D^{\bar{j}_1} \cdots D^{\bar{j}_m}
  \right\} \nonumber \\ 
 & ~~~~~~~~~+ a^{(n)}_2 
  \left(\partial_{\bar{i}}\Phi D^{\bar{l}}
   + \delta_{il} \partial_{\bar{j}}\Phi D^{\bar{j}}
   \right).
  \label{lhs}
\end{align}
On the other hand, the right hand side of (\ref{rec-An}) is
calculated as
\begin{align}
 \left[\partial_{\bar{i}}, A_{n-1}\right]
  &= \sum_{m=2}^{n-1} a^{(n-1)}_m 
  \left[ \partial_{\bar{i}}, 
   \partial_{\bar{j}_1}\Phi \cdots \partial_{\bar{j}_{m-1}}\Phi
   D^{\bar{j}_1} \cdots D^{\bar{j}_m} D^{\bar{l}}
  \right]
  \nonumber \\
% &= \sum_{m=2}^{n-1}c^{(n-1)}_m
%  \left\{
%   m \partial_{\bar{j}_1}\Phi \cdots \partial_{\bar{j}_{m-1}}\Phi
%   \partial_{\bar{i}}\Phi
%   D^{\bar{j}_1} \cdots D^{\bar{j}_{m-1}}D^{\bar{l}}
%   + \delta_{il} \partial_{\bar{j}_1}\Phi \cdots \partial_{\bar{j}_m}\Phi
%   D^{\bar{j}_1} \cdots D^{\bar{j}_m}
%   \right. \nonumber \\
% & + (m-1)^2 \partial_{\bar{j}_1}\Phi \cdots \partial_{\bar{j}_{m-2}}\Phi
%   \partial_{\bar{i}}\Phi
%   D^{\bar{j}_1} \cdots D^{\bar{j}_{m-2}}D^{\bar{l}}
%   \nonumber \\
% & \left.
%   + (m-1)\delta_{il} \partial_{\bar{j}_1}\Phi \cdots 
%   \partial_{\bar{j}_{m-1}}\Phi
%   D^{\bar{j}_1} \cdots D^{\bar{j}_{m-1}}
%  \right\}
% \nonumber \\
 &= \sum_{m=2}^{n-1}
  \left(a^{(n-1)}_m + m a^{(n-1)}_{m+1}\right)
  \nonumber \\
 & ~~~~~~\times 
  \left(
   m \partial_{\bar{j}_1}\Phi \cdots \partial_{\bar{j}_{m-1}}\Phi
   \partial_{\bar{i}}\Phi
   D^{\bar{j}_1} \cdots D^{\bar{j}_{m-1}}D^{\bar{l}}
   + \delta_{il} \partial_{\bar{j}_1}\Phi \cdots \partial_{\bar{j}_m}\Phi
   D^{\bar{j}_1} \cdots D^{\bar{j}_m}
  \right)
  \nonumber \\
 & ~~~~~~+ a^{(n-1)}_2 \left(\partial_{\bar{i}}\Phi D^{\bar{l}}
	       + \delta_{il} \partial_{\bar{j}} D^{\bar{j}}\right).
 \label{rhs}
\end{align}
Equating (\ref{lhs}) with (\ref{rhs}), we find
\begin{align}
 a^{(n)}_2 = a^{(n-1)}_2 = \cdots = a^{(2)}_2 =1,
 \label{a2}
\end{align}
and the following recursion relation
\begin{align}
 \label{rec-c}
 a^{(n)}_m = a^{(n-1)}_{m-1} + (m-1) a^{(n-1)}_m.
\end{align}
To solve this equation, we introduce a generating function
\begin{align}
 \alpha_m (t) \equiv \sum_{n=m}^\infty t^n a^{(n)}_m, \label{f-def}
\end{align}
for $m \geq 2$.
Then the relation (\ref{rec-c}) is written as
\begin{align}
 \alpha_m (t) = t\left[\alpha_{m-1}(t) + (m-1) \alpha_m (t)
 \right]. \label{f-rec}
\end{align}
This is solved as
\begin{align}
 \alpha_{m}(t) &= \frac{t}{1-(m-1)t} \alpha_{m-1}(t) \nonumber \\ &=
 t^{m-2} \prod_{n=2}^{m-1}\frac{1}{1-nt} \times 
 \alpha_2 (t). \label{fm-f2}
\end{align}
Since $\alpha_2(t)$ is easily calculated from (\ref{a2}) as
\begin{align}
 \alpha_2 (t) = \sum_{n=2}^\infty t^n a^{(n)}_2 = \sum_{n=2}^\infty t^n
  = \frac{t^2}{1-t}, \label{f2}
\end{align}
$\alpha_m (t)$ is determined as
\begin{align}
 \alpha_m(t) = t^m \prod_{n=1}^{m-1}\frac{1}{1-nt}
  = \frac{\Gamma(1-m+\frac{1}{t})}{\Gamma(1+\frac{1}{t})}, 
  \qquad (m\geq 2). 
 \label{fm}
\end{align}
The function $\alpha_m (t)$ actually coincides with the generating
function for the Stirling numbers of the second kind $S(n,k)$, and
$a^{(n)}_m$ is related to $S(n,k)$ as
\begin{align}
 a^{(n)}_{m} = S(n-1, m-1). \label{2ndS}
\end{align}

Summarizing the above calculations, $L_{\bar{z}^l}$ becomes
\begin{align}
 L_{\bar{z}^l} &= \bar{z}^l + \hbar D^{\bar{l}}
  + \sum_{n=2}^\infty \hbar^n \sum_{m=2}^n a^{(n)}_m
  \partial_{\bar{j}_1}\Phi \cdots \partial_{\bar{j}_{m-1}}\Phi
  D^{\bar{j}_1} \cdots D^{\bar{j}_{m-1}} D^{\bar{l}}
  \nonumber \\
 &= \bar{z}^l + \hbar D^{\bar{l}} 
  + \sum_{m=2}^\infty 
  \left( \sum_{n=m}^\infty \hbar^n a^{(n)}_m \right)
  \partial_{\bar{j}_1}\Phi \cdots \partial_{\bar{j}_{m-1}}\Phi
  D^{\bar{j}_1} \cdots D^{\bar{j}_{m-1}} D^{\bar{l}}
  \nonumber \\
  &= \bar{z}^l
   + \sum_{m=1}^\infty \alpha_m(\hbar)
  \partial_{\bar{j}_1}\Phi \cdots \partial_{\bar{j}_{m-1}}\Phi
  D^{\bar{j}_1} \cdots D^{\bar{j}_{m-1}} D^{\bar{l}}. \label{L_z}
\end{align}
Here we defined $\alpha_1(t)=t$. Similarly, it can be shown that the
right star product by $z^l$, $R_{z^l} f = f * z^l$ is expressed as
\begin{align} 
 R_{z^l} &= z^l + \hbar D^l
  + \sum_{n=2}^\infty \hbar^n \sum_{m=2}^n a^{(n)}_m
  \partial_{j_1}\Phi \cdots \partial_{j_{m-1}}\Phi
  D^{j_1} \cdots D^{j_{m-1}} D^l
  \nonumber \\
&= z^l 
  + \sum_{m=1}^\infty \alpha_m(\hbar)
  \partial_{j_1}\Phi \cdots \partial_{j_{m-1}}\Phi
  D^{j_1} \cdots D^{j_{m-1}} D^l, \label{R_z}
\end{align}
where $D^i = g^{i\bar{j}}\partial_{\bar{j}}$.\\

{} From the theorem \ref{theo1}, proposition \ref{prop1},
(\ref{L_z}) and (\ref{R_z}), we obtain the following theorem.
\begin{thm} \label{3theorem}
A star product with separation of variables for
$\mathbb{C}P^N$ with the K\"ahler
potential
$
 \Phi = \ln \left(1+|z|^2\right)
$ is given by 
\begin{align}
f * g = L_f g = R_g f.
\end{align}
Here differential operators $L_f$ and $R_g$ 
are determined by the differential operators $L_{\bar{z}}$ and $R_z$ 
whose expressions are given in (\ref{L_z}) and (\ref{R_z})
through the relation 
(\ref{L_f}) and (\ref{R_f}), respectively.
\end{thm}

We can now calculate the star products among $z^i$ and $\bar{z}^i$,
\begin{align}
 z^i*z^j &= z^iz^j, \label{z-z} \\
 z^i*\bar{z}^j &= z^i\bar{z}^j, \label{z-barz} \\
 \bar{z}^i*\bar{z}^j &= \bar{z}^i\bar{z}^j, \label{barz-barz} \\
 \bar{z}^i*z^j &= 
 \bar{z}^iz^j + \hbar \delta_{ij} (1+|z|^2)
  {}_2F_1\left(1, 1; 1-1/\hbar; -|z|^2\right) \nonumber \\
 & ~~~~+\frac{\hbar}{1-\hbar} \bar{z}^i z^j (1+|z|^2) 
 {}_2F_1 \left(1, 2; 2-1/\hbar; -|z|^2\right), \label{barz-z}
\end{align}
where ${}_2F_1$ is the Gauss hypergeometric function.
Here we used the following equation
\begin{align}
 D^{\bar{j}_1} \cdots D^{\bar{j}_m}z^i
  &= (m-1)! (1+|z|^2)^m  \nonumber \\
  & ~~~~\times \Bigg[
   \sum_{k=1}^{m}\delta_{i j_k} \bar{z}^{j_1}\cdots 
   \hat{\bar{z}}^{j_k}\cdots \bar{z}^{j_{m}}
   + m z^i \bar{z}^{j_1}\cdots \bar{z}^{j_{m}} \Bigg], \label{D-z}
\end{align}
where the hat over a term means that it is to be omitted from the
product.

There are several ways of having deformation quantization by a reduction from
higher dimensional manifolds. 
Within the framework of Karabegov's method
for K\"ahler manifolds, a general reduction procedure of deformation
quantizations with separation 
of variables was considered in \cite{Karabegov2008}.
When the standard star product
with separation of variables corresponding to a K\"ahler potential 
$\rho = \psi(z, \bar{z}) u\bar{u}$, where $z$ and $u$ are holomorphic
coordinates, is given, the reduction procedure eliminates the variables 
$u, \bar{u}$ and produces
the standard star product with separation of variables corresponding
to the K\"ahler potential $\ln |\psi|$.
This reduction procedure can be applied to the case of the reduction from 
$\mathbb{C}^{N+1}\backslash \{0\}$ to $\mathbb{C}P^N$ and gives the same
star product as the one used in this article.
On the other hand, in \cite{Bordemann}, a star product on
$\mathbb{C}P^N$ was constructed 
by performing the phase space reduction from 
$\mathbb{C}^{N+1} \backslash \{0\}$.  
The expression of their star product, denoted as $*_B$, for functions $f$
and $g$ on $\mathbb{C}P^N$ is given
\begin{align}
 & f *_B g \nonumber \\ 
 & = fg + \sum_{m=1}^\infty \hbar^m
 \sum_{s=1}^m \sum_{k=1}^s 
 \frac{k^{m-1} (-1)^{m-k}}{s! (s-k)! (k-1)!}  
 \left(|\zeta|^2\right)^s
 \frac{\partial^s f}{\partial \bar{\zeta}^{A_1}\cdots\bar{\zeta}^{A_s}}
 \frac{\partial^s g}{\partial \zeta^{A_1}\cdots\zeta^{A_s}},
\end{align} 
where $\zeta^{A_i}, \bar{\zeta}^{A_j}$ are the homogeneous coordinates.
This is also the star product with separation of variables, and thus
(\ref{z-z})-(\ref{barz-barz}) hold trivially under $*_B$ product.
$\bar{z}^i *_B z^j$ is calculated as
\begin{align}
 \bar{z}^i *_B z^j =& \bar{z}^i z^j
 + \hbar \delta_{ij} (1+|z|^2) \tilde{F}_1(-|z|^2)
 + \hbar \bar{z}^i z^j (1+|z|^2) \tilde{F}_2(-|z|^2),
\end{align}
where $z^i = \zeta^i/\zeta^0, \bar{z}^i = \bar{\zeta}^i/\bar{\zeta}^0$,
and 
\begin{align}
 \tilde{F}_1 (-|z|^2) \equiv& 
  \sum_{m=0}^\infty \sum_{s=0}^m \sum_{k=1}^{s+1}
 \frac{\hbar^m s! k^m (-1)^{m+1-k}}{(s+1-k)! (k-1)!}(1+|z|^2)^s, \\
 \tilde{F}_2(-|z|^2) \equiv& 
  \sum_{m=0}^\infty \sum_{s=0}^m \sum_{k=1}^{s+1}
 \frac{\hbar^m (s+1)! k^m (-1)^{m+1-k}}{(s+1-k)! (k-1)!}(1+|z|^2)^s.
\end{align} 
We can show that $\tilde{F}_1 (-|z|^2)$ satisfies the hypergeometric
equation and the boundary conditions for 
${}_2F_1(1, 1; 1-1/\hbar; -|z|^2)$, and thus 
$\tilde{F}_1 (-|z|^2) = {}_2F_1(1, 1; 1-1/\hbar; -|z|^2)$.  
Similarly, 
$\tilde{F}_2 (-|z|^2) = {}_2F_1(1, 2; 2-1/\hbar; -|z|^2)/(1-\hbar)$ 
can be also shown. 
Therefore it turns out $\bar{z}^i * z^j = \bar{z}^i *_B z^j$.  
These facts lead to $f*g = f*_B g$.
Namely, this calculation shows that the star product constructed by
Karabegov's method coincides with the star product $*_B$ in 
\cite{Bordemann}.
As far as we know, the origin of this coincidence of the star products
obtained by these different methods is not apparent at this time.

%%%%%%%%%%%%%%%%%%%%%%%%%%%%%%
\section{Fock representation}
\label{FOCK}
%%%%%%%%%%%%%%%%%%%%%%%%%%%%
%\setcounter{equation}{0}

The left star product by $\partial_i \Phi$ and the right star
product by $\partial_{\bar{i}} \Phi$ are respectively written as
\begin{align}
 L_{\partial_i \Phi} &= \hbar \partial_i + \partial_i \Phi
 = \hbar e^{-\Phi/\hbar} \partial_i e^{\Phi/\hbar}, 
 \label{LdPhi}\\
 R_{\partial_{\bar{i}} \Phi} 
 &= \hbar \partial_{\bar{i}} + \partial_{\bar{i}} \Phi
 = \hbar e^{-\Phi/\hbar} \partial_{\bar{i}} e^{\Phi/\hbar}.
 \label{RdPhi}
\end{align} 
{}From the definition of the star product given in the previous section,
we easily find 
\begin{align}
 \partial_i \Phi*z^j - z^j*\partial_i \Phi &= \hbar\delta_{ij}, 
 \qquad 
 z^i * z^j - z^j * z^i = 0, \qquad
 \partial_i \Phi * \partial_j \Phi 
 - \partial_j \Phi * \partial_i \Phi = 0, 
 \label{comm-rel-1} \\
 \bar{z}^i*\partial_{\bar{j}}\Phi - \partial_{\bar{j}}\Phi*\bar{z}^i
 &= \hbar\delta_{ij}, \qquad
 \bar{z}^i * \bar{z}^j - \bar{z}^j * \bar{z}^i = 0, \qquad
 \partial_{\bar{i}} \Phi * \partial_{\bar{j}} \Phi 
 - \partial_{\bar{j}} \Phi * \partial_{\bar{i}} \Phi = 0.
 \label{comm-rel-2}
\end{align}
Hence, $\{z^i, \partial_j \Phi ~|~ i, j=1, 2, \cdots, N \}$ and 
$\{\bar{z}^i, \partial_{\bar{j}} \Phi ~|~ i, j=1, 2, \cdots, N \}$ 
constitute $2N$ sets of the creation-annihilation operators
under the star product. But, it is noted that operators in 
$\{z^i, \partial_j \Phi\}$ does not commute with ones in 
$\{\bar{z}^i, \partial_{\bar{j}} \Phi\}$, e.g., 
$z^i * \bar{z}^j - \bar{z}^j * z^i \neq 0$.

Here, we would like to construct the Fock representation of the star
product. First we show that $e^{-\Phi/\hbar}=(1+|z|^2)^{-1/\hbar}$ is
the vacuum projection. $e^{-\Phi/\hbar}$ is annihilated by the left
star product of $\partial_i \Phi$ and $\bar{z}^i$,
\begin{align}
 \partial_i \Phi * e^{-\Phi/\hbar} &= L_{\partial_i \Phi} e^{-\Phi/\hbar}
 = \hbar e^{-\Phi/\hbar} \partial_i e^{\Phi/\hbar} e^{-\Phi/\hbar}
 = 0, \label{dPhi-e}\\
 \bar{z}^i * e^{-\Phi/\hbar} &= L_{\bar{z}^i} e^{-\Phi/\hbar} 
 \nonumber \\ 
 &= \left(\bar{z}^i + \sum_{m=1}^\infty
 \alpha_m(\hbar) \partial_{\bar{j}_1}\Phi \cdots \partial_{\bar{j}_{m-1}}\Phi  
 D^{\bar{j_1}} \cdots D^{\bar{j}_{m-1}} \right) e^{-\Phi/\hbar}
 \nonumber \\
 &= 0. \label{z-e}
\end{align} 
Here the following equation is used
\begin{align}
 D^{\bar{j_1}} \cdots D^{\bar{j}_{m}} e^{-\Phi/\hbar}
 &= D^{\bar{j_1}} \cdots D^{\bar{j}_{m}} (1+|z|^2)^{-1/\hbar}
 \nonumber \\
 &= (-1)^m \frac{\Gamma(1+1/\hbar)}{\Gamma(1-m+1/\hbar)}
 z^{j_1} \cdots z^{j_{m-1}} (1+|z|^2)^{m-1/\hbar}. \label{D-e}
\end{align}
Similarly, it is shown that $e^{-\Phi/\hbar}$ is annihilated by the right
star product of the $\partial_{\bar{i}} \Phi$ and $z^i$,
\begin{equation}
 e^{-\Phi/\hbar} * \partial_{\bar{i}} \Phi
  = e^{-\Phi/\hbar} * z^i = 0. 
\end{equation}

Next, we show that $e^{-\Phi/\hbar}$ satisfies the relation 
\begin{equation}
e^{-\Phi/\hbar} * f(z, \bar{z}) = e^{-\Phi/\hbar} f(0, \bar{z})
 \label{vacuum1}
\end{equation}
for a function $f(z, \bar{z})$ such that $f(z, \bar{w})$ can be expanded
as Taylor series with respect to $z^i$ and $\bar{w}^j$, respectively. 
To show
the relation, we note that the differential operator $R_{z^i}$
corresponding to the right product of $z^i$ contains only partial
derivatives by $\bar{z}^j$, and thus commutes with $z^k$. Moreover,
$R_{z^i}$ annihilates $e^{-\Phi/\hbar}$, 
$R_{z^i} e^{-\Phi/\hbar} = e^{-\Phi/\hbar} * z^i =0$ as
mentioned above. From these, the relation (\ref{vacuum1}) is shown as
\begin{align}
 e^{-\Phi/\hbar} * f(z, \bar{z}) &= R_f e^{-\Phi/\hbar} \nonumber \\
 &= \sum_{k_1, \dots, k_N=0}^\infty \frac{1}{k_1! \cdots k_N!}
 \partial_1^{k_1} \cdots \partial_N^{k_N} f(z, \bar{z})
 \prod_{m=1}^{N} \left(R_{z^m} - z^m \right)^{k_m} e^{-\Phi/\hbar}
 \nonumber \\
 &= \sum_{k_1, \dots, k_N=0}^\infty \frac{1}{k_1! \cdots k_N!}
 \partial_1^{k_1} \cdots \partial_N^{k_N} f(z, \bar{z})
 \prod_{m=1}^{N} \left(- z^m \right)^{k_m} e^{-\Phi/\hbar}
 \nonumber \\
 &= e^{-\Phi/\hbar} f(0, \bar{z}). \label{e-f}
\end{align}
Similarly, the following equation holds
\begin{equation}
 f(z, \bar{z}) * e^{-\Phi/\hbar} = f(z, 0) e^{-\Phi/\hbar}.
  \label{vacuum2}
\end{equation}
As a specific case of the equation (\ref{vacuum1}), the idempotency of
$e^{-\Phi/\hbar}$ is obtained,
\begin{align}
 e^{-\Phi(z, \bar{z})/\hbar} * e^{-\Phi(z, \bar{z})/\hbar} 
 = e^{-\Phi(z, \bar{z})/\hbar} e^{-\Phi(0, \bar{z})/\hbar}
 = e^{-\Phi(z, \bar{z})/\hbar}
 \label{idem},
\end{align}
where $\Phi(0, \bar{z})=0$ is used.

By using the relations (\ref{vacuum1}) and (\ref{vacuum2}), it is
possible to calculate explicitly star products containing
$e^{-\Phi/\hbar}$ as follows,
\begin{align}
 e^{-\Phi/\hbar} * \left(\partial_{i_1}\Phi(z, \bar{z}) \cdots
 \partial_{i_n}\Phi(z, \bar{z}) \right) 
 &= e^{-\Phi/\hbar} \left(\partial_{i_1}\Phi(0, \bar{z}) \cdots
 \partial_{i_n}\Phi(0, \bar{z}) \right) \nonumber \\
 &= \bar{z}^{i_1} \cdots \bar{z}^{i_n} e^{-\Phi/\hbar} \nonumber \\
 &= e^{-\Phi/\hbar} * \bar{z}^{i_1} * \cdots * \bar{z}^{i_n}, \\
 \left(\partial_{\bar{i}_1}\Phi(z, \bar{z}) \cdots
 \partial_{\bar{i}_n}\Phi(z, \bar{z}) \right) * e^{-\Phi/\hbar}
 &= z^{i_1} \cdots z^{i_n} e^{-\Phi/\hbar} \nonumber \\
 &= z^{i_1} * \cdots * z^{i_n} * e^{-\Phi/\hbar}. \label{dPhi-e-z-e}
\end{align}

We then consider a class of functions
\begin{align} \label{M_ij}
 M_{i_1 \cdots i_m ;  j_1 \cdots j_n} 
 &= \frac{z^{i_1}\cdots z^{i_m} \bar{z}^{j_1} \cdots \bar{z}^{j_n}}
 {\sqrt{m!n!\alpha_m(\hbar)\alpha_n(\hbar)}}
 e^{-\Phi/\hbar},
\end{align}
where $\alpha_n(\hbar)$ is defined in (\ref{fm}). 
$M_{i_1 \cdots i_m ;  j_1 \cdots j_n}$ is totally symmetric under
permutations of $i$'s and $j$'s, respectively.
It is also useful to represent this function as  
\begin{align}
 M_{i_1 \cdots i_m ;  j_1 \cdots j_n}
 &= \frac{1}{\sqrt{m!n!\alpha_m(\hbar)\alpha_n(\hbar)}}
 z^{i_1} * \cdots * z^{i_m} * e^{-\Phi/\hbar} *
 \left( \partial_{j_1}\Phi \cdots \partial_{j_n}\Phi \right) 
 \nonumber \\
 &= \frac{1}{\hbar^n} \sqrt{\frac{\alpha_n(\hbar)}{m!n!\alpha_m(\hbar)}}
 z^{i_1} * \cdots * z^{i_m} * e^{-\Phi/\hbar} *
 \partial_{j_1}\Phi * \cdots * \partial_{j_n}\Phi \nonumber \\ \label{M}
 &= \frac{1}{\sqrt{m!n! \alpha_m(\hbar) \alpha_n(\hbar)}}
 \left(\partial_{{\bar i}_1}\Phi \cdots \partial_{{\bar i}_m}\Phi
 \right) * e^{-\Phi/\hbar} *{\bar z}^{j_1} * \cdots * {\bar z}^{j_n} 
 \nonumber \\ 
 &= \frac{1}{\hbar^m}\sqrt{\frac{\alpha_m(\hbar)}{m!n!\alpha_n(\hbar)}}
 \partial_{{\bar i}_1}\Phi * \cdots * \partial_{{\bar i}_m}\Phi 
 * e^{-\Phi/\hbar} *{\bar z}^{j_1} * \cdots * {\bar z}^{j_n}.
\end{align}
In the second equality, we used the following relation,
\begin{align}
 \partial_{j_1}\Phi * \cdots * \partial_{j_n}\Phi 
 &= L_{\partial_{j_1}\Phi} \cdots L_{\partial_{j_n}\Phi} 1 \nonumber \\
 &= \left(\hbar e^{-\Phi/\hbar} \partial_{j_1} e^{\Phi/\hbar} \right)
 \cdots 
 \left(\hbar e^{-\Phi/\hbar} \partial_{j_n} e^{\Phi/\hbar} \right) 1
 \nonumber \\
 &= \frac{\hbar^n \Gamma(1+1/\hbar)}{\Gamma(1-n+1/\hbar)}
 \bar{z}^{j_1} \cdots \bar{z}^{j_n} (1+|z|^2)^{-n} \nonumber \\
 &= \frac{\hbar^n \Gamma(1+1/\hbar)}{\Gamma(1-n+1/\hbar)}
 \partial_{j_1}\Phi \cdots \partial_{j_n}\Phi. \label{dPhi*s}
\end{align}
By using the commutation relations (\ref{comm-rel-1}) and the fact that
$e^{-\Phi/\hbar}$ is the vacuum projection, it can be shown that these
functions form a closed algebra:
\begin{align}
 M_{i_1 \cdots i_m; j_1 \cdots j_n} * M_{k_1 \cdots k_r; l_1 \cdots l_s}
% &= \frac{\Gamma(1-n+1/\hbar)}{\hbar^n \Gamma(1+1/\hbar)}
% \left( z^{i_1} * \cdots * z^{i_m} * e^{-\Phi/\hbar} *
% \partial_{j_1}\Phi * \cdots * \partial_{j_n}\Phi \right)
% \nonumber \\
% & ~~~~ * \left(z^{k_1} * \cdots * z^{k_r} * e^{-\Phi/\hbar} *
% \bar{z}^{l_1} * \cdots * \bar{z}^{l_s} \right)
% \nonumber \\
 &= \delta_{nr} \delta^{k_1 \cdots k_n}_{j_1 \cdots j_n}
 M_{i_1 \cdots i_m; l_1 \cdots l_s}, \label{M-alg}
\end{align}
where $\delta^{k_1 \cdots k_n}_{j_1 \cdots j_n}$ is defined as
\begin{equation}
 \delta^{k_1 \cdots k_n}_{j_1 \cdots j_n}
  = \frac{1}{n!} \left[\delta^{k_1}_{j_1} \cdots \delta^{k_n}_{j_n} 
  + \mbox{permutations of } (j_1, \cdots, j_n)\right]. 
  \label{multi-delta}
\end{equation}

At last, the following theorem is obtained.
\begin{thm}
Let 
$\displaystyle 
{\cal M} = \left\{1,\sum_{i,j} a_{ij} M_{i ; j}  \right\} $
be a set of linear combinations of  
$M_{i_1 \cdots i_m ; j_1 \cdots j_n}$
defined by (\ref{M_ij}) in $\mathbb{C}P^N$, 
where $i, j$ are multi-index
of $i=(i_1, \cdots , i_m )$ and $j=(j_1 , \cdots , j_n)$
and $a_{ij} \in {\mathbb C}$.
%Each $M_{i_1 \cdots i_m ; j_1 \cdots j_n}$ has relations through 
%the creating operators $\{ z^i , \partial_j \Phi \}$
%and annihilation operators $\{ \bar{z}^i , \partial_{\bar{j}} \Phi \}$
%as (\ref{M-relation1})-(\ref{M-relation2}), each other.
Then ${\cal M}$ is a ring whose 
multiplication is defined by
the star product in Theorem \ref{3theorem},
and its algebra is given by (\ref{M-alg}).
\end{thm}

Further, the star products between $M_{i_1 \cdots i_m; j_1 \cdots j_n}$
and one of $z^k, \partial_k \Phi, \bar{z}^k$ and
$\partial_{\bar{k}}\Phi$ are calculated as follows,
\begin{align} \label{M-relation1}
 z^k * M_{i_1 \cdots i_m; j_1 \cdots j_n} 
 &= \sqrt{\frac{m+1}{-m+1/\hbar}} M_{k i_1 \cdots i_m; j_1 \cdots j_n}, \\
 \partial_k \Phi * M_{i_1 \cdots i_m; j_1 \cdots j_n}
 &= \hbar \sqrt{\frac{-m+1+1/\hbar}{m}} 
 \sum_{l=1}^m \delta_{k i_l} 
 M_{i_1 \cdots \hat{i_l} \cdots i_m; j_1 \cdots j_n},  \\
 \bar{z}^k * M_{i_1 \cdots i_m; j_1 \cdots j_n}
 &= \frac{1}{\sqrt{m(-m+1+1/\hbar)}}
 \sum_{l=1}^m \delta_{k i_l} 
 M_{i_1 \cdots \hat{i_l} \cdots i_m; j_1 \cdots j_n}, \\
 \partial_{\bar{k}} \Phi * M_{i_1 \cdots i_m; j_1 \cdots j_n}
 &= \hbar \sqrt{(m+1)(-m+1/\hbar)}
 M_{k i_1 \cdots i_m; j_1 \cdots j_n}, \\
 M_{i_1 \cdots i_m; j_1 \cdots j_n} * z^k
 &= \frac{1}{\sqrt{n(-n+1+1/\hbar)}} 
 \sum_{l=1}^n \delta_{k j_l} 
 M_{i_1 \cdots i_m; j_1 \cdots \hat{j_l} \cdots j_n}, \\
 M_{i_1 \cdots i_m; j_1 \cdots j_n} * \partial_k \Phi
 &= \hbar \sqrt{(n+1)(-n+1/\hbar)} M_{i_1 \cdots i_m; j_1 \cdots j_n k},
 \\
 M_{i_1 \cdots i_m; j_1 \cdots j_n} * \bar{z}^k
 &= \sqrt{\frac{n+1}{-n+1/\hbar}} M_{i_1 \cdots i_m; j_1 \cdots j_n k}, \\
 M_{i_1 \cdots i_m; j_1 \cdots j_n} * \partial_{\bar{k}} \Phi
 &= \hbar \sqrt{\frac{-n+1+1/\hbar}{n}}
 \sum_{l=1}^n \delta_{k j_l}
 M_{i_1 \cdots i_m; j_1 \cdots \hat{j_l} \cdots j_n}.
\label{M-relation2}
\end{align}

%%%%%%%%%%%%%%%%%%%%%%%%%%%%%%%%
\section{The case of $\mathbb{C}\mathbf{H^N}$}
\label{CHN}
%%%%%%%%%%%%%%%%%%%%%%%%%%%%%%%%%%%

The K\"ahler potential of $\mathbb{C}H^N$ is given by
\begin{align}
 \Phi = - \ln \left(1-|z|^2\right).
\end{align}
%where $z_i (i=1,\cdots,N)$ are the inhomogeneous coordinates and $|z|^2 = \sum_{k=1}^N z^k \bar{z}^k$.
The metric $g_{i\bar{j}}$ and the inverse metric $g^{\bar{i}j}$ are
defined by 
\begin{align}
% ds^2 &= 2g_{i\bar{j}}dz^id\bar{z}^j, \\
 g_{i\bar{j}} &= \partial_i \partial_{\bar{j}} \Phi
  = \frac{(1-|z|^2)\delta_{ij}+\bar{z}^i z^j}{(1-|z|^2)^2}, \\
 g^{\bar{i}j} &= (1-|z|^2)\left(\delta_{ij}-\bar{z}^i z^j\right).
\end{align}
Then we find the following relations similar to
(\ref{par-phi})-(\ref{par-a-D});
\begin{align}
 \partial_{\bar{i_1}} \partial_{\bar{i_2}} \cdots 
  \partial_{\bar{i}_n} \Phi 
  =&~ (n-1)! ~\partial_{\bar{i}_1}\Phi
  \partial_{\bar{i}_2}\Phi \cdots \partial_{\bar{i}_n}\Phi, \\
 \left[\partial_{\bar{i}}, D^{\bar{j}}\right]
  =&~ - \partial_{\bar{i}}\Phi D^{\bar{j}} 
  - \delta_{ij}\partial_{\bar{k}}\Phi D^{\bar{k}}, \\
% \left[\partial_{\bar{i}}, a_{\bar{j}_1 \bar{j}_2 \cdots \bar{j}_n}
%  D^{\bar{j}_1} D^{\bar{j}_2} \cdots D^{\bar{j}_n} \right]
% =&~ - n \partial_{\bar{i}}\Phi ~a_{\bar{j}_1 \cdots \bar{j}_n}
% D^{\bar{j}_1} \cdots D^{\bar{j}_n} \nonumber \\
% &~ - n a_{\bar{i}\bar{j}_1 \cdots \bar{j}_{n-1}} \partial_{\bar{k}} \Phi
%  D^{\bar{k}} D^{\bar{j}_1} \cdots D^{\bar{j}_{n-1}} 
%  \nonumber \\
% &~ - n(n-1)a_{\bar{i} \bar{j}_1 \cdots \bar{j}_{n-1}}
%  D^{\bar{j}_1} \cdots D^{\bar{j}_{n-1}}. \nonumber \\
%%%%%%%%%%%%%%%%
 \left[\partial_{\bar{i}}, ~c_{\bar{j}_1 \bar{j}_2 \cdots \bar{j}_n}
 D^{\bar{j}_1} D^{\bar{j}_2} \cdots D^{\bar{j}_n} \right]
  =&~ \partial_{\bar{i}} c_{\bar{j}_1 \bar{j}_2 \cdots \bar{j}_n}
 D^{\bar{j}_1} D^{\bar{j}_2} \cdots D^{\bar{j}_n}
 - n c_{\bar{j}_1 \cdots \bar{j}_n} \partial_{\bar{i}}\Phi 
 D^{\bar{j}_1} \cdots D^{\bar{j}_n} \nonumber \\
 & \hspace{-30mm} 
 - n c_{\bar{i}\bar{j}_1 \cdots \bar{j}_{n-1}} \partial_{\bar{k}} \Phi
  D^{\bar{k}} D^{\bar{j}_1} \cdots D^{\bar{j}_{n-1}} 
%  \nonumber \\
% &~ 
 - n(n-1) c_{\bar{i} \bar{j}_1 \cdots \bar{j}_{n-1}}
  D^{\bar{j}_1} \cdots D^{\bar{j}_{n-1}}. 
%%%%%%%%%%%%%
\end{align}

The operator $L_{\bar{z}^l}$ is
expanded as a power series of the noncommutative parameter $\hbar$,
\begin{align}
 L_{\bar{z}^l} = \bar{z}^l + \hbar D^{\bar{l}}
  + \sum_{n=2}^\infty \hbar^n B_n. \label{Lzh_ch}
\end{align}
We assume that $B_n$ has the following form,
\begin{align}
 \label{Bn}
 B_n = \sum_{m=2}^n (-1)^{n-1} b^{(n)}_m
  \partial_{\bar{j}_1}\Phi \cdots \partial_{\bar{j}_{m-1}}\Phi
  D^{\bar{j}_1} \cdots D^{\bar{j}_{m-1}} D^{\bar{l}}.
\end{align}
The factor $(-1)^{n-1}$ in the front of the coefficient $b^{(n)}_m$ is
introduced for convenience.

Requiring $\left[L_{\bar{z}^l}, \partial_{\bar{i}}\Phi + \hbar
\partial_{\bar{i}} \right] = 0$, it is found that $b^{(n)}_m$ should
satisfy similar relations to (\ref{a2}) and (\ref{rec-c}),
\begin{align}
 b^{(n)}_2 &= b^{(n-1)}_2 = \cdots = b^{(2)}_2 =1, \nonumber \\
 b^{(n)}_m &= b^{(n-1)}_{m-1} + (m-1) b^{(n-1)}_m.
\end{align}
Hence $b^{(n)}_m$ coincides with $a^{(n)}_m$, and 
we obtain the explicit representation of the star product with
separation of variables on $\mathbb{C}H^N$,
\begin{align}
 L_{\bar{z}^l} &= \bar{z}^l + \hbar D^{\bar{l}}
  + \sum_{n=2}^\infty \hbar^n \sum_{m=2}^n 
 (-1)^{n-1} b^{(n)}_m
  \partial_{\bar{j}_1}\Phi \cdots \partial_{\bar{j}_{m-1}}\Phi
  D^{\bar{j}_1} \cdots D^{\bar{j}_{m-1}} D^{\bar{l}}
  \nonumber \\
  &= \bar{z}^l
   + \sum_{m=1}^\infty (-1)^{m-1} \beta_m(\hbar)
  \partial_{\bar{j}_1}\Phi \cdots \partial_{\bar{j}_{m-1}}\Phi
  D^{\bar{j}_1} \cdots D^{\bar{j}_{m-1}} D^{\bar{l}}, 
 \label{Lz-ch} \\
 R_{z^l} &= z^l + \hbar D^l
  + \sum_{n=2}^\infty \hbar^n \sum_{m=2}^n (-1)^{n-1} b^{(n)}_m
  \partial_{j_1}\Phi \cdots \partial_{j_{m-1}}\Phi
  D^{j_1} \cdots D^{j_{m-1}} D^l
  \nonumber \\
&= z^l 
  + \sum_{m=1}^\infty (-1)^{m-1} \beta_m(\hbar)
  \partial_{j_1}\Phi \cdots \partial_{j_{m-1}}\Phi
  D^{j_1} \cdots D^{j_{m-1}} D^l,
 \label{Rz-ch}
\end{align}
with
\begin{align}
\beta_n(t)=(-1)^n \alpha_n(-t)= \frac{\Gamma(1/t)}{\Gamma(n+1/t)}.
\label{beta-n}
\end{align}

{} From the theorem \ref{theo1} and proposition \ref{prop1}, we obtain the
following theorem.
\begin{thm}
A star product with separation of variables for
$\mathbb{C}H^N$ with the K\"ahler
potential
$
 \Phi = -\ln \left(1-|z|^2\right)
$ is given by 
\begin{align}
f * g = L_f g = R_g f.
\end{align}
Here differential operators $L_f$ and $R_g$ are determined through the
relation (\ref{L_f}) and (\ref{R_f}), respectively, by the differential
operators $L_{\bar{z}}$ and $R_z$ whose expressions are given in
 (\ref{Lz-ch}) and (\ref{Rz-ch}).

\end{thm}

Using the representations of the star product, we can calculate the star
products among $z^i$ and $\bar{z}^i$,
\begin{align}
 z^i*z^j =& z^iz^j, \\
 z^i*\bar{z}^j =& z^i\bar{z}^j, \\
 \bar{z}^i*\bar{z}^j =& \bar{z}^i\bar{z}^j, \\
 \bar{z}^i*z^j =& 
  \bar{z}^iz^j + \hbar \delta_{ij} (1-|z|^2) 
 {}_2F_1\left(1, 1; 1 + 1/\hbar; |z|^2 \right)\nonumber \\
 & - \frac{\hbar}{1+\hbar} \bar{z}^i z^j (1-|z|^2)
 {}_2F_1 \left(1, 2; 2+1/\hbar; |z|^2\right).
\end{align}
Here the following equation similar to (\ref{D-z}) is used,
\begin{align}
 D^{\bar{j}_1} \cdots D^{\bar{j}_{m}} z^i
  &= (-)^{m-1} (m-1)! (1-|z|^2)^m  \nonumber \\
  & ~~~~\times \Bigg[
   \sum_{k=1}^{m}\delta_{i j_k} \bar{z}^{j_1} \cdots 
   \hat{\bar{z}}^{j_k}\cdots \bar{z}^{j_m}
   - m z^i \bar{z}^{j_1} \cdots \bar{z}^{j_m} \Bigg].
\end{align}

As in the case of $\mathbb{C}P^N$, $\{z^i, \partial_j \Phi\}$ and
$\{\bar{z}^i, \partial_{\bar{j}} \Phi\}$ satisfy the commutation 
relations for the creation-annihilation operators.
Also $e^{-\Phi/\hbar}$ is the vacuum projection operator,
\begin{align}
 \partial_i \Phi* e^{-\Phi/\hbar} &= 0, \\
 \bar{z}^i * e^{-\Phi/\hbar} &= 0, \\
 e^{-\Phi/\hbar} * \partial_{\bar{i}} \Phi &=0, \\
 e^{-\Phi/\hbar} * z^i &= 0,
\end{align} 
and 
\begin{align}
 e^{-\Phi/\hbar} * e^{-\Phi/\hbar} &
 =  e^{-\Phi/\hbar}.
\end{align}
Here we used the following relation which is corresponding to 
(\ref{D-e}) in the case of $\mathbb{C}P^N$,
\begin{align}
 D^{\bar{j_1}} \cdots D^{\bar{j}_m} e^{-\Phi/\hbar}
 &= D^{\bar{j_1}} \cdots D^{\bar{j}_m} (1-|z|^2)^{1/\hbar}
 \nonumber \\
 &= (-1)^m \frac{\Gamma(1/\hbar + m)}{\Gamma(1/\hbar)}
 z^{j_1} \cdots z^{j_m} (1-|z|^2)^{1/\hbar + m}.
\end{align}

As in the case of $\mathbb{C}P^N$, we consider a class of functions
\begin{align}
N_{i_1 \cdots i_m ; j_1 \cdots j_n} 
 &= \frac{z^{i_1} \cdots z^{i_m} \bar{z}^{j_1} \cdots \bar{z}^{j_n}}
 {\sqrt{m!n! \beta_m(\hbar) \beta_n(\hbar)}} 
 e^{-\Phi/\hbar} \nonumber \\ 
 &= \frac{1}{\sqrt{m!n! \beta_m(\hbar) \beta_n(\hbar)}}
 z^{i_1} * \cdots * z^{i_m} * e^{-\Phi/\hbar} 
 * \left(\partial_{j_1}\Phi \cdots  \partial_{j_n}\Phi \right) 
 \nonumber \\ 
 &= \frac{1}{\hbar^n}\sqrt{\frac{\beta_n(\hbar)}{m!n!\beta_m(\hbar)}}
 z^{i_1} * \cdots * z^{i_m} * e^{-\Phi/\hbar} *
 \partial_{j_1}\Phi * \cdots * \partial_{j_n}\Phi \nonumber \\
&= \frac{1}{\sqrt{m!n! \beta_m(\hbar) \beta_n(\hbar)}}
 \left(\partial_{{\bar i}_1}\Phi \cdots \partial_{{\bar i}_m}\Phi
 \right) * e^{-\Phi/\hbar} *{\bar z}^{j_1} * \cdots * {\bar z}^{j_n} 
 \nonumber \\ 
 &= \frac{1}{\hbar^m}\sqrt{\frac{\beta_m(\hbar)}{m!n!\beta_n(\hbar)}}
 \partial_{{\bar i}_1}\Phi * \cdots * \partial_{{\bar i}_m}\Phi 
 * e^{-\Phi/\hbar} *{\bar z}^{j_1} * \cdots * {\bar z}^{j_n}.
\end{align}
$N_{i_1 \cdots i_m ;  j_1 \cdots j_n}$ is totally symmetric under
permutations of $i$'s and $j$'s, respectively.
Then we can show that these functions form a closed algebra 
\begin{align}
N_{i_1 \cdots i_m ; j_1 \cdots j_n} * 
 N_{k_1 \cdots k_r ; l_1 \cdots l_s} 
 &= \delta_{n r} \delta^{k_1 \cdots k_n}_{j_1 \cdots j_n}
 N_{i_1 \cdots i_m ; l_1
 \cdots l_s}. 
\end{align}
Moreover, the star products between $N_{i_1 \cdots i_m; j_1 \cdots j_n}$
and one of $z^k, \partial_k \Phi, \bar{z}^k$ and
$\partial_{\bar{k}}\Phi$ are calculated as follows,
\begin{align}
 z^k * N_{i_1 \cdots i_m; j_1 \cdots j_n} 
 &= \sqrt{\frac{m+1}{m+1/\hbar}} N_{k i_1 \cdots i_m; j_1 \cdots j_n}, \\
 \partial_k \Phi * N_{i_1 \cdots i_m; j_1 \cdots j_n}
 &= \hbar \sqrt{\frac{m-1+1/\hbar}{m}} 
 \sum_{l=1}^m \delta_{k i_l} 
 N_{i_1 \cdots \hat{i_l} \cdots i_m; j_1 \cdots j_n},  \\
 \bar{z}^k * N_{i_1 \cdots i_m; j_1 \cdots j_n}
 &= \frac{1}{\sqrt{m(m-1+1/\hbar)}}
 \sum_{l=1}^m \delta_{k i_l} 
 N_{i_1 \cdots \hat{i_l} \cdots i_m; j_1 \cdots j_n}, \\
 \partial_{\bar{k}} \Phi * N_{i_1 \cdots i_m; j_1 \cdots j_n}
 &= \hbar \sqrt{(m+1)(m+1/\hbar)}
 N_{k i_1 \cdots i_m; j_1 \cdots j_n}, \\
 N_{i_1 \cdots i_m; j_1 \cdots j_n} * z^k
 &= \frac{1}{\sqrt{n(n-1+1/\hbar)}} 
 \sum_{l=1}^n \delta_{k j_l} 
 N_{i_1 \cdots i_m; j_1 \cdots \hat{j_l} \cdots j_n}, \\
 N_{i_1 \cdots i_m; j_1 \cdots j_n} * \partial_k \Phi
 &= \hbar \sqrt{(n+1)(n+1/\hbar)} N_{i_1 \cdots i_m; j_1 \cdots j_n k},
 \\
 N_{i_1 \cdots i_m; j_1 \cdots j_n} * \bar{z}^k
 &= \sqrt{\frac{n+1}{n+1/\hbar}} N_{i_1 \cdots i_m; j_1 \cdots j_n k}, \\
 N_{i_1 \cdots i_m; j_1 \cdots j_n} * \partial_{\bar{k}} \Phi
 &= \hbar \sqrt{\frac{n-1+1/\hbar}{n}}
 \sum_{l=1}^n \delta_{k j_l}
 N_{i_1 \cdots i_m; j_1 \cdots \hat{j_l} \cdots j_n}.
\end{align}

%%%%%%%%%%%%%%%%%
\section{Summary and discussion}
\label{Summary}
%%%%%%%%%%%%%%%%%

  In this paper, we obtained explicit expressions of star
products in ${\mathbb C}P^N$ and ${\mathbb C}H^N$ by using the
deformation quantization with separation of variables proposed by
Karabegov. In this quantization method, a star product by a function
is represented by a formal series of differential operators, which is
obtained as the solution of an infinite system of differential
equations. We gave the explicit solutions of the equations in the case of
${\mathbb C}P^N$ and ${\mathbb C}H^N$. The operators corresponding
to the left (right) star multiplications of functions are determined as
the power series of $\hbar$ in which each term contains the Stirling
numbers of the second kind, the K\"ahler potentials of the manifolds,
and the differential operators.

  We also constructed the Fock representations of the star products by
using the fact that $\{z^i, \partial_j \Phi\}$ and $\{z^{\bar{i}},
\partial_{\bar{j}} \Phi\}$ constitute $2N$ sets of the
creation-annihilation operators under the star product. We first
identified the function $e^{-\Phi/\hbar}$ corresponding to the vacuum
projection. Then we considered the functions which are derived by
multiplying polynomials of $z^i$ and $z^{\bar{i}}$ on
$e^{-\Phi/\hbar}$, and showed that these functions form the closed algebra
under the star product.

Now, we have three comments.  
Firstly, the operator $L_f$ of the left star
multiplication by a function $f$ which is given in Section \ref{CPN} for
$\mathbb{C}P^N$ and in Section \ref{CHN} for $\mathbb{C}H^N$,
respectively, can  be represented by using the covariant derivatives. 
To this end, we show that $L_f$ on these manifolds has the following form,
\begin{align}
 L_f &= \sum_{n=0}^\infty c_n (\hbar) 
 g_{j_1 \bar{k}_1} \cdots g_{j_n \bar{k}_n} 
 \left(D^{j_1} \cdots D^{j_n} f\right)
 D^{\bar{k}_1} \cdots D^{\bar{k}_n}.
 \label{Lf-cov}
\end{align}
The coefficient $c_n(\hbar)$ is determined by the condition 
$[L_f, \hbar \partial_{\bar{i}} + \partial_{\bar{i}} \Phi] =0$.
For the case of $\mathbb{C}P^N$, this condition becomes
\begin{align}
 [L_f, \hbar \partial_{\bar{i}} + \partial_{\bar{i}} \Phi] &=
 \sum_{n=1}^\infty 
 \left[ n (1-\hbar(n-1)) c_n(\hbar) - \hbar c_{n-1} (\hbar) \right]
 \nonumber \\
 & \hspace{10mm} \times
 g_{l \bar{i}} g_{j_1 \bar{k}_1} \cdots g_{j_{n-1} \bar{k}_{n-1}}
 \left(D^l D^{j_1} \cdots D^{j_{n-1}} f \right)
 D^{\bar{k}_1} \cdots D^{\bar{k}_{n-1}}
 =0.
\end{align}
By solving the recursion relation, 
$n (1-\hbar(n-1)) c_n(\hbar) - \hbar c_{n-1} (\hbar) = 0$, 
under the initial condition $c_0 =1$,
$c_n(\hbar)$ is obtained as
\begin{align}
 c_n (\hbar) &= \frac{\Gamma(1-n+1/\hbar)}{n! \Gamma(1+1/\hbar)}
 = \frac{\alpha_n (\hbar)}{n!},
\end{align}
where $\alpha_n (\hbar)$ in given in (\ref{fm}).
The first two terms in the power series of $\hbar$ of $L_f$ are
calculated,
\begin{align}
 L_f &\equiv f + \hbar g_{j\bar{k}} \left(D^j f\right) D^{\bar{k}} 
 ~~({\rm mod} ~\hbar^2).
\end{align}
Similarly, the operator $L_f$ on $\mathbb{C}H^N$ can be represented in
the form of (\ref{Lf-cov}) with $c_n(\hbar) = \beta_n (\hbar)/n!$ where 
$\beta_n(\hbar)$ is defined in (\ref{beta-n}).

The expression of $L_f$ (\ref{Lf-cov}) can be rewritten by the use of the
covariant derivatives on the manifolds. Non-vanishing components of the
Christoffel symbols on a K\"ahler manifolds are only $\Gamma^i_{jk}$ and 
$\Gamma^{\bar{i}}_{\bar{j}\bar{k}}$. Hence, for scalars $f$ and $g$
\begin{align}
 g^{j_1 \bar{k}_1} \cdots g^{j_n \bar{k}_n} 
 \nabla_{\bar{k}_1} \cdots \nabla_{\bar{k}_n} f &= 
  g^{j_1 \bar{k}_1}\nabla_{\bar{k}_1}
 \left(g^{j_2 \bar{k}_2} \cdots g^{j_n \bar{k}_n} 
 \nabla_{\bar{k}_2} \cdots \nabla_{\bar{k}_n} f\right) \nonumber \\
 &= g^{j_1 \bar{k}_1}\partial_{\bar{k}_1}
 \left(g^{j_2 \bar{k}_2} \cdots g^{j_n \bar{k}_n} 
 \nabla_{\bar{k}_2} \cdots \nabla_{\bar{k}_n} f\right) \nonumber \\
 &= D^{j_1} \left(g^{j_2 \bar{k}_2} \cdots g^{j_n \bar{k}_n} 
 \nabla_{\bar{k}_2} \cdots \nabla_{\bar{k}_n} f\right) \nonumber \\
 &= D^{j_1} \cdots D^{j_n} f, \\
 g^{\bar{j}_1 k_1} \cdots g^{\bar{j}_n k_n} 
 \nabla_{k_1} \cdots \nabla_{k_n} g
 &= D^{\bar{j}_1} \cdots D^{\bar{j}_n} g.
\end{align}
Using these relations. $L_f g$ becomes
\begin{align}
 L_f g &= f * g = \sum_{n=0}^\infty c_n (\hbar) 
 g^{\bar{j}_1 k_1} \cdots g^{\bar{j}_n k_n} 
 \left(\nabla_{\bar{j}_1} \cdots \nabla_{\bar{j}_n} f\right)
 \left(\nabla_{k_1} \cdots \nabla_{k_n} g\right).
 \label{Lfg-cov}
\end{align}

In this article, we treat $\hbar$ as a formal parameter.
Now we consider the specific case of $\hbar=1/L ~(L \in \mathbb{N})$ and
the star product in a function space ${\cal M}_L$ spanned by
$$
\frac{z^{i_1} \cdots z^{i_m} \bar{z}^{j_1} \cdots \bar{z}^{j_n}}
{(1+|z|^2)^L}, \qquad (m, n \leq L).
$$
In this case, the series in (\ref{Lf-cov}) terminates at $n=L$, because
$$
 D^{j_1} \cdots D^{j_{L+1}} f  = 0, \qquad
 D^{\bar{k}_1} \cdots D^{\bar{k}_{L+1}} g  = 0,
$$
where $f, g \in {\cal M}_L$. Then, the expression of the star product
coincides with the one in \cite{Balachandran}.

Secondly, let us try to extend the covariant expression of $L_f$ 
(\ref{Lfg-cov}) to locally symmetric K\"ahler manifolds, 
$\nabla_\mu {R_{\nu\rho\sigma}}^\lambda = 0$.
We assume the following form of $L_f$,
\begin{align}
 L_f g &= \sum_{n=0}^\infty T_n^{\bar{j}_1 \cdots \bar{j}_n, k_1 \cdots k_n}
 \left(\nabla_{\bar{j}_1} \cdots \nabla_{\bar{j}_n} f\right)
 \left(\nabla_{k_1} \cdots \nabla_{k_n} g \right).
\end{align}
Here $g$ is a scalar function and 
$T_n^{\bar{j}_1 \cdots \bar{j}_n, k_1 \cdots k_n}$ 
is a covariantly constant tensor, $\nabla T_n = 0$, 
and completely symmetric under permutations of
$\bar{j}$'s and $k$'s, respectively.  
Requiring $[L_f, \partial_{\bar{i}}\Phi + \hbar \partial_{\bar{i}}]=0$, 
recursion relations for 
$T_n^{\bar{j}_1 \cdots \bar{j}_n, k_1 \cdots k_n}$ are derived,
\begin{align}
 \Big[
 & nT_n^{\bar{j}_1 \cdots \bar{j}_n, k_1 \cdots k_n} g_{k_n \bar{i}}
  -\hbar T_{n-1}^{\bar{j}_1 \cdots \bar{j}_{n-1}, k_1 \cdots k_{n-1}} 
 \delta_{\bar{i}}^{\bar{j}_n} 
  \nonumber \\
 & ~~~~ -\hbar \frac{n(n-1)}{2}
 T_n^{\bar{j}_1 \cdots \bar{j}_n, k_1 \cdots k_{n-2} p q}
 {R_{\bar{i}pq}}^{k_{n-1}}
 \Big]
 \left(\nabla_{\bar{j}_1} \cdots \nabla_{\bar{j}_n} f \right)
 \left(\nabla_{k_1} \cdots \nabla_{k_{n-1}} g \right) = 0.
\end{align} 
Since the recursion relations include only the metric and the
Riemann tensor, 
$T_n^{\bar{j}_1 \cdots \bar{j}_n, k_1 \cdots k_n}$ is determined as 
a function of these quantities and satisfies $\nabla T_n = 0$. 
Further, the recursion relations are simplified in the case of 
$\mathbb{C}P^N$. Because 
$R_{i\bar{j}k\bar{l}} =
-g_{i\bar{j}}g_{k\bar{l}}-g_{i\bar{l}}g_{k\bar{j}}$
on $\mathbb{C}P^N$ with the metric (\ref{metric}), it can be shown that
$L_f$ has the covariant form (\ref{Lfg-cov}).

Thirdly, we consider relations between star products
on different patches.
Let $ \bigcup U_i $ $( U_i = \{ [\zeta^0 : \zeta^1 : \cdots : \zeta^N ] 
(\zeta^i \neq 0 )\})$ be an open covering of ${\mathbb C}P^N$,
where $\zeta^k$ is a homogeneous coordinate.
We define an inhomogeneous coordinate $z^k = \frac{\zeta^k}{\zeta^0}$
in $U_0$ and  $w_1= \frac{\zeta^0}{\zeta^1}, 
w^k = \frac{\zeta^k}{\zeta^1} ~(k \ge 2)$
in $U_1$ .
Consider the mapping from $U_0$ to $U_1$.
The transformation between these inhomogeneous coordinates
is given by 
\begin{align}
w^1 & =\frac{1}{z^1}, ~w^k= \frac{z^k}{z^1} ~(k \ge 2).
\end{align}
Under this transformation, the K\"ahler potential
is changed as
\begin{align}
\Phi (z) &= \ln (1+|z|^2)=\ln (1+|w|^2)-\ln w^1-\ln \bar{w}^1 .
\end{align}
{}From the lemma 3 in \cite{Karabegov1996},
it is found that the star product does not change under the transformation.
Similarly, the form of the star product is invariant
under mappings between other patches.

\vspace{10mm}
%%%%%%%%%%%%%%%%%%%%%%%%%%%%%%%%%%%%%%%%%%%%%

\noindent {\bf Acknowledgement}\\ A.S. is supported by KAKENHI
No.23540117 (Grant-in-Aid for Scientific Research (C)). H.U. is
supported by KAKENHI No.21740197 (Grant-in-Aid for Young Scientists
(B)). We should like to thank an anonymous referee for his helpful comments.

\vspace{10mm}
\section*{Appendix}
\renewcommand{\theequation}{A.\arabic{equation}}

Star products on $\mathbb{C}P^1$ have been well studied. In
the appendix, we summarize the results in the case of $\mathbb{C}P^1$
for convenience.

In the case of $\mathbb{C}P^1$, $L_{\bar{z}}$ and $R_{z}$ are given by
\begin{align}
 L_{\bar{z}} &= \bar{z} + 
  \sum_{m=1}^\infty \alpha_m(\hbar)
  \left(\bar{\partial}\Phi\right)^{m-1} \bar{D}^m, \\
 R_{z} &= z + \sum_{m=1}^\infty \alpha_m(\hbar) 
	 \left(\partial \Phi\right)^{m-1} D^m,
\end{align}
where $\Phi = \ln (1+z\bar{z})$ is the K\"ahler potential of
$\mathbb{C}P^1$, $\bar{D}=g^{\bar{z}z}\partial=(1+z\bar{z})^2 \partial$,
and $D = g^{z\bar{z}}\bar{\partial}=(1+z\bar{z})^2 \bar{\partial}$.
$\alpha_m(\hbar)$ is defined in (\ref{fm}).
The star products among $z$ and $\bar{z}$ become
\begin{align}
 z * z &= z^2, \\
 z * \bar{z} &= |z|^2, \\
 \bar{z} * \bar{z} &= \bar{z}^2, \\
 \bar{z} * z &= |z|^2 + \hbar(1+|z|^2)^2 ~
 _2F_1 (1, 2; 1-1/\hbar; -|z|^2).
\end{align}

Under the star product, $\partial \Phi$ and $z$ ($\bar{z}$ and
$\bar{\partial} \Phi$) satisfy the commutation relations of the
creation-annihilation operators, respectively,
\begin{align}
 \partial \Phi * z - z * \partial \Phi &= \hbar, \\
 \bar{z} * \bar{\partial} \Phi - \bar{\partial} \Phi * \bar{z} &= \hbar. 
\end{align}

The function $e^{-\Phi/\hbar}=(1+|z|^2)^{-1/\hbar}$ is corresponding to
the vacuum projection,
\begin{align}
 \partial \Phi * e^{-\Phi/\hbar} &= \bar{z} * e^{-\Phi/\hbar}
 = e^{-\Phi/\hbar} * \bar{\partial} \Phi = e^{-\Phi/\hbar} * z =0, \\
 e^{-\Phi/\hbar} * e^{-\Phi/\hbar} &= e^{-\Phi/\hbar}.
\end{align}

The following functions 
\begin{align}
 M_{mn} &= 
 \frac{z^m \bar{z}^n}{\sqrt{m!n!\alpha_m(\hbar)\alpha_n(\hbar)}}
 e^{-\Phi/\hbar}
\end{align}
form a closed algebra,
\begin{align}
 M_{mn} * M_{kl} &= \delta_{nk} M_{ml}.
\end{align}
These formulae coincide with the ones of the fuzzy sphere \cite{Madore,
Alexanian} when $\hbar = 1/L ~(L \in \mathbb{N})$, as mentioned in
Section \ref{Summary}.

%%%%! The bibliography is not in alphabetical order.

%%%%%%%%%%%%%%%%%%%%%%%%%%%%%%%

\end{document}